\documentclass[modern]{aastex62}
\usepackage{amsmath}
\usepackage{mathtools}
\usepackage{color}
\usepackage{hyperref}
\usepackage[normalem]{ulem}

\newcommand{\dd}{{\mathrm d}}
\newcommand{\rmsub}[1]{_\mathrm{#1}}
\newcommand{\bmu}{\boldsymbol{\mu}}
\newcommand{\beps}{\boldsymbol{\varepsilon}}
\newcommand{\blam}{\boldsymbol{\Lambda}}
\newcommand{\vx}[1]{{\bf {#1}}}
\newcommand{\vxhat}[1]{{\bf {\hat{#1}}}}
\newcommand{\pkgname}{\texttt{amlc}}

\graphicspath{{./}{figures/}}

\shortauthors{Tchernyshyov}

\begin{document}

\title{Analytic marginalization of absorption line continua}

\author[0000-0003-0789-9939]{Kirill Tchernyshyov}
\affiliation{Department of Physics and Astronomy, Johns Hopkins University, 3400 N. Charles Street, Baltimore, MD 21218, USA}
\correspondingauthor{Kirill Tchernyshyov}
\email{ktcherny@gmail.com}

\begin{abstract}
Absorption line spectroscopy is a powerful way of measuring properties of stars and the interstellar medium.
Absorption spectra are often analyzed manually, an approach that limits reproducibility and which cannot practically be applied to modern datasets consisting of thousands or even millions of spectra.
Simultaneous probabilistic modeling of absorption features and continuum shape is a promising approach for automating this analysis.
Existing implementations of this approach use numerical methods such as Markov chain Monte Carlo (MCMC) to marginalize over the continuum parameters.
Numerical marginalization over large numbers of continuum parameters is too slow for exploratory analysis, can increase the dimensionality of an inference problem beyond the capacity of simple MCMC samplers, and is in general impractical for the analysis of large datasets.
When continua are parameterized as linear functions such as polynomials or splines, it is possible to reduce continuum parameter marginalization to an integral over a multivariate normal distribution, which has a known closed form.
In addition to speeding up probabilistic modeling, analytic marginalization makes it trivial to marginalize over continuum parameterizations and to combine continuum description marginalization with optimization for absorption line parameters.
These new possibilities allow automatic, probabilistically justified continuum placement in analyses of large spectroscopic datasets.
We compare the accuracy to within which absorption line parameters can be recovered using different continuum placement methods and find that marginalization is in many cases an improvement over other methods.
We implement analytic marginalization over linear continuum parameters in the open-source package \pkgname.
\end{abstract}


\section{Introduction}
\label{sec:introduction}
Absorption lines contain information on the composition and properties of interstellar matter (ISM) and stellar atmospheres.
To extract this information, it is necessary to decompose the spectrum into absorption features and the intrinsic flux, typically referred to as the continuum, produced by the illuminating background source towards which the absorption is seen.
The most common way of doing this separation has been manually finding regions in a spectrum that do not contain absorption features, fitting a function to these regions, and using this function to interpolate over the absorption features.
Given the longevity and popularity of this approach, it is clear that it can produce acceptable results.
It does, however, have two important weaknesses.
The first is that every spectrum must be examined and interacted with by a human.
This cannot efficiently be done for datasets containing thousands or even millions of spectra.
The second is that it is unlikely that the absorption parameter estimator this procedure implicitly defines uses data efficiently.
There is variance between analyses done by different humans as well as between analyses done by the same human at different times.
If there is a subset of analysts whose estimates are the most accurate and precise, then the estimates of the rest are using the available data inefficiently.

An alternative approach is to infer absorption line and continuum parameters simultaneously.
To improve the accuracy of the inferred absorption line parameters, it can be useful to marginalize over, rather than fit for, the continuum parameters.
This has been done in packages meant for the analysis of absorption lines from both the ISM (\texttt{BayesVP}, \citealt{Liang:2018kq}) and stellar atmospheres (\texttt{Starfish}, \citealt{2015ApJ...812..128C}, and \texttt{sick}, \citealt{2016ApJS..223....8C}).
In these packages, continuum parameter marginalization is done numerically, using Markov chain Monte Carlo (MCMC).
As the authors of two of these packages point out, including large numbers of continuum parameters in MCMC sampling leads to long convergence and autocorrelation times.
To keep the number of continuum parameters low, these packages either do not support (\texttt{BayesVP}) or advise against (\texttt{sick}) including continuum parameters when simultaneously analyzing multiple spectral segments.

Even when there are few continuum parameters, using MCMC to analyze a modern spectroscopic dataset consisting of thousands or even millions of spectra will be computationally demanding.
Probablistic analyses of comparable numbers of photometric observations \citep[e.g.][]{2015ApJ...810...25G,2016ApJ...826..104G} require months or years of computation time.
A single spectrum contains orders of magnitude more data points than a single set of photometric observations, which comes with an at least linearly proportional increase in required computation time.

In the packages discussed above and in much of the absorption line analysis literature, the continuum is assumed to be a low order polynomial or spline.
While these are non-linear functions of wavelength, they are linear functions of the polynomial or spline coefficients.
This linearity means that if some additional assumptions hold, it is possible to marginalize over these coefficients analytically.

Analytic marginalization has several advantages over numerical marginalization.
First, it can speed up MCMC-based inference for absorption line parameters by reducing the dimensionality of the problem.
Second, the ability to evaluate the continuum parameter-marginalized likelihood function and its gradient makes it possible to do optimization, rather than sampling, for the absorption parameters while still keeping the robustness provided by marginalizing over continuum parameters.
Finally, it makes marginalization over different possible continuum parameterizations computationally trivial---simply sum together continuum-marginalized likelihoods that assume different parameterizations.
Marginalization over parameterizations allows an even greater degree of automation and systematization of absorption line inference.

The assumptions required for this particular form of analytic marginalization are: that the continuum can be expressed as a linear function (not necessarily a polynomial or spline); that the priors on the parameters of this linear function are either improper uniform or (multivariate) normal; and that residuals from the model are normally distributed.
If these assumptions hold, then, given a model for the absorption, the posterior probability distribution function of the continuum parameters is itself a multivariate normal distribution.
The result of marginalizing over the continuum parameters is simply an update to the covariance matrix of the model residuals.
Conceptually, when a set of absorption line parameters is specified, the continuum parameters can be treated as additive, rather than multiplicative, linear nuisance parameters.
An explanation of marginalization over additive linear nuisance parameters in an astronomical context is given in \citet{2017RNAAS...1a...7L}.
This approach to marginalizing over multiplicative linear nuisance parameters has already been used for several astronomical applications, for example for analyzing sparsely sampled radial velocity measurements \citep{2017ApJ...837...20P}.

Models for absorption line spectra have features, such as the presence of a line spread function (LSF), which should be accounted for to more efficiently compute marginalized likelihoods and likelihood gradients.
In this work, we derive expressions for these quantities that account for these features.
This derivation is given in Section \ref{sec:assumptions-and-formalism}.
We have created a package, \pkgname\footnote{\pkgname{} is available at \url{https://github.com/ktchrn/amlc}.}, for evaluating these expressions.
The package is described in Appendix \ref{sec:package-and-demos}.
The performance of continuum parameter and parameterization marginalization is explored in Section \ref{sec:test-cases}.
We discuss strengths and weaknesses of this method in Section \ref{sec:discussion} and conclude in Section \ref{sec:conclusion}.

\section{Assumptions and formalism}
\label{sec:assumptions-and-formalism}
We assume the following model for a spectrum $\vx{y}$ given parameters $\theta$, $\vx{m}$, and $\vx{b}$:
\begin{align}
\label{eqn:basic-model-expanded}
\vx{y}(\theta) &= \vx{L} \left( \vx{d}(\theta) \odot \left(\bmu_m(\theta) + \sum_{i=1}^P \vx{a}_{m,i}\, m_i  \right)
 +\bmu_b(\theta) + \sum_{i=1}^Q \vx{a}_{b,i} \,b_i \right) + \beps.
\end{align}
The background source emits a continuum, which is expressed as the sum of a possibly non-linear term, $\bmu_m(\theta)$, and a linear combination of basis elements $\vx{a}_{m,i}$ with coefficients $m_i$.
Intervening matter absorbs part of this continuum with transmittance function $\vx{d}(\theta)$.
The absorption happens independently at each wavelength.
This is indicated by the elementwise product $\odot$ between the transmittance and continuum.
Foregrounds, such as sky lines or instrumental artifacts are, like the continuum, expressed as the sum of a possibly non-linear term, $\bmu_b(\theta)$, and a linear combination of basis elements $\vx{a}_{b,i}$ with coefficients $b_i$.
The resulting spectrum is convolved with an LSF $\vx{L}$ and observed.
$\beps$ are the residuals between the observed $\vx{y}$ and the LSF-convolved spectrum and are assumed to be normally distributed with mean zero and covariance matrix $\vx{K}$.
The length of the observed spectrum is $M$, the length of the pre-LSF model spectrum is $N$, the number of continuum basis elements is $P$, and the number of foreground basis elements is $Q$.

Collecting the multiplicative (continuum) and additive (foreground) basis elements $\vx{a}_{m,i}$ and $\vx{a}_{b,i}$ into matrices $\vx{A_m}$ and $\vx{A_b}$ and converting the transmittance vector $\vx{d}(\theta)$ into the diagonal matrix $\vx{D}_\theta \equiv \text{diag} \left( \vx{d}(\theta) \right) $,
\begin{align}
  \label{eqn:basic-model-compact}
  \vx{y}  &= \vx{L} \left( \bmu_b(\theta) + \vx{A_b} \vx{b}
  + \vx{D}_\theta \left( \bmu_m(\theta) + \vx{A_m} \vx{m} \right) \right) + \beps\\
  &\equiv \vx{L} \left( \bmu_b(\theta) + \vx{D}_\theta \bmu_m(\theta) + \vx{B} \vx{c} \right) + \beps.
\end{align}
In the second expression, $\vx{B}$ and $\vx{c}$ are defined as:
\begin{align}
  \label{eqn:c-definitions}
  \vx{B} &= \begin{bmatrix}
  \vx{D}_\theta \vx{A_m} & \vx{A_b}
  \end{bmatrix}
  &\vx{c} = \begin{bmatrix}
  \vx{m}\\
  \vx{b}
  \end{bmatrix}.
\end{align}
We consider two possible priors for the nuisance parameter vector $\vx{c}$, a multivariate normal distribution with mean zero and covariance matrix $\blam$ and an improper uniform distribution:
\begin{align}
  \label{eqn:priors}
  p_n(\vx{c}) = \mathcal{N}\left({\bf 0}, \blam \right) \text{ (normal) \hspace{0.2in} and \hspace{0.2in}} p_u(\vx{c}) = \prod_{i=1}^{P+Q} Z_i^{-1} \text{ (uniform)},
\end{align}
where $Z_i$ can be any positive real number.

\subsection{Conditional probability of the nuisance parameters}
\label{sec:conditionals}
For both priors, the conditional distribution of $\vx{c}$ at fixed $\theta$ is proportional to a multivariate normal distribution.
The mean $\vxhat{c}$ of this normal distribution is
\begin{align}
  \label{eqn:conditional-c-mean}
  \vxhat{c}_{n/u} &= \vx{C}^{-1}_{n/u} \vx{B}^T \vx{L}^T \vx{K}^{-1} \vx{r},
\end{align}
where $\vx{r}$ is the vector of residuals
\begin{align}
  \label{eqn:def-r}
  \vx{r} &= \vx{y} - \vx{L} \left( \bmu_b(\theta) + \vx{D}_\theta \bmu_m(\theta) \right)
\end{align}
and $\vx{C}_{n/u}$ is
\begin{align}
  \vx{C}_n &= \blam^{-1} + \vx{B}^T \vx{L}^T \vx{K}^{-1} \vx{L} \vx{B},
\end{align}
if the prior on $\vx{c}$ is normal, and
\begin{align}
  \vx{C}_u &= \vx{B}^T \vx{L}^T \vx{K}^{-1} \vx{L} \vx{B}
\end{align}
if the prior on $\vx{c}$ is uniform.
The covariance matrix of the conditional distribution of $\vx{c}$ is $\vx{C}_{n/u}^{-1}$.

The conditional distribution of $\vx{c}$ can be used for visualization and predictive checks.
The mean of the conditional distribution is also its mode, so $\vx{L} \vx{B} \vxhat{c}$ is the best-fit model for $\vx{y}$ at a given value of $\theta$.
Samples drawn from the conditional distribution of $\vx{c}$ can be used to visualize the effect and extent of nuisance parameter variation.

\subsection{Marginal likelihood}
Assuming the normal prior $p_n(\vx{c})$, marginalizing over $\vx{c}$ following e.g. \citet{2017RNAAS...1a...7L} or \citet{Rasmussen:2006vz} gives
\begin{align}
  \label{eqn:proper-prior-marginal}
  p_n(\vx{y} \vert \theta, \vx{L}, \vx{B}, \vx{K}, \blam) &=
   \int_{-\infty}^{+\infty} p(\vx{y} \vert \vx{c}, \theta, \vx{L}, \vx{B}, \vx{K}, \blam) p_n(\bf c) \,\dd\vx{c}\\
  &= (2\pi)^{-\frac{M}{2}} \det(\vx{K})^{-\frac{1}{2}} \det(\blam)^{-\frac{1}{2}} \det(\vx{C}_n)^{-\frac{1}{2}}
  \exp \left[ -\frac{1}{2}  \vx{r}^T \vx{K}^{-1} \left( \vx{r} - \vxhat{r}_n \right) \right],
\end{align}
where
\begin{align}
  \vxhat{r}_{n/u} &= \vx{L}\vx{B} \vxhat{c}_{n/u}.
\end{align}
If we instead assume the improper prior $p_u(\vx{c})$,
\begin{align}
  \label{eqn:improper-prior-marginal}
  p_u(\vx{y} \vert \theta, \vx{L}, \vx{B}, \vx{K}) &=
  \int_{-\infty}^{+\infty} p(\vx{y} \vert \vx{c}, \theta, \vx{L}, \vx{B}, \vx{K}) p_u(\bf c) \, \dd \vx{c}\\
  &= \left( \prod_{i=1}^{P+Q} Z_i^{-1} \right) (2\pi)^{-\frac{M - (P+Q)}{2}} \det(\vx{K})^{-\frac{1}{2}} \det(\vx{C}_u)^{-\frac{1}{2}}
  \exp \left[ -\frac{1}{2}  \vx{r}^T \vx{K}^{-1} \left( \vx{r} - \vxhat{r}_u \right) \right].
\end{align}
The marginal likelihood $p_u$ will be proper if $\vx{C}_u$ is positive definite, which will be the case when $\vx{L}\vx{B}$ is full rank and $M \geq P+Q$.
The marginal likelihood $p_n$ is always proper because $\vx{C}_n$ is always positive definite.
$\vx{C}_n$ is always positive definite because $\blam^{-1}$ is always positive definite and $\vx{B}^T \vx{L}^T \vx{K}^{-1} \vx{L} \vx{B}$ is always at least positive semi-definite.

\subsection{Gradients}
We give expressions for the gradients of $\log(p_n)$ and $\log(p_u)$ with respect to $\vx{d}(\theta)$, $\bmu_b(\theta)$, and $\bmu_m(\theta)$.
The gradient of $\log(p)$ with respect to the parameters $\theta$ can be obtained by evaluating each of these gradients, computing the Jacobians of $\vx{d}(\theta)$, $\bmu_b(\theta)$, and $\bmu_m(\theta)$ with respect to $\theta$, and applying the chain rule.

The gradient of $\log(p)$ with respect to ${\bf d}(\theta)$ is
\begin{align}
  \label{eqn:grad-logp-wrt-dtheta}
  \begin{split}
  \nabla \log(p)&({\bf d(\theta)}) = \left( \vx{L}^T \vx{K}^{-1}\left(\vx{r} - \vxhat{r}_{n/u} \right) \right)
  \odot \left( \vx{B}' \vxhat{c} + \bmu_m \right) \\
  &- \frac{1}{2} \left( \left( \vx{C}_{n/u}^{-1} \vx{B}'^T \right) \odot \left(\vx{B}^T \vx{L}^T \vx{K}^{-1} \vx{L}  \right)
  + \left( \vx{C}_{n/u}^{-1} \vx{B}^T \vx{L}^T \vx{K}^{-1} \vx{L} \right) \odot \vx{B}'^T
    \right) \vx{1},
  \end{split}
\end{align}
where $\vx{1}$ is a column vector of ones of length $P+Q$.
$\vx{B'}$ is the sum of derivatives of $\vx{B}$ with respect to each element of $\vx{d}(\theta)$:
\begin{align}
\vx{B'} &= \sum_{i=1}^N \frac{\partial \vx{B}}{\partial d_i(\theta)} \\
&= \sum_{i=1}^N
\begin{bmatrix}
\vx{J}^{i, i} \vx{A_m} & 0 \times \vx{A_b}
\end{bmatrix} \\
&=
\begin{bmatrix}
\vx{A_m} & \vx{0}
\end{bmatrix},
\end{align}
where $\vx{J}^{i, i}$ is a square matrix whose $(i,i)$-th entry is 1 and whose other entries are all 0.
The first row of Equation \ref{eqn:grad-logp-wrt-dtheta} is the gradient of the argument of the exponentials in Equations \ref{eqn:proper-prior-marginal} and \ref{eqn:improper-prior-marginal}.
The second row is the gradient of $\log\left(\det \left( \vx{C}_{n/u}\right) \right)$.

The gradient of $\log(p)$ with respect to $\bmu_m(\theta)$ is
\begin{align}
  \label{eqn:grad-logp-wrt-mum}
  \nabla \log(p)(\bmu_m(\theta)) &= \vx{D}_\theta \vx{L}^T \vx{K}^{-1} \left( \vx{r} - \vxhat{r}_{n/u} \right)
\end{align}
and the gradient of $\log(p)$ with respect to $\bmu_b(\theta)$ is
\begin{align}
  \label{eqn:grad-logp-wrt-mub}
  \nabla \log(p)(\bmu_b(\theta)) &= \vx{L}^T \vx{K}^{-1} \left( \vx{r} - \vxhat{r}_{n/u} \right).
\end{align}

\section{Practical test cases}
\label{sec:test-cases}
Does marginalization over continuum parameters, analytic or numerical, have benefits beyond having a clear probablistic justification?
We consider two metrics: how marginalizing over, instead of fitting for, continuum parameters and parameterizations affects the error to within which absorption line parameters can be measured; and how marginalizing analytically, instead of numerically, affects the speed of MCMC-based inference.

For the error metric, we consider two possible cases: one in which the continuum parameterization is known and one in which it is necessary to choose a continuum parameterization from a set of possibilites.
The suggested method is marginalizing over continuum parameters, in the first case, and marginalizing over continuum parameters and parameterizations, in the second.
When the parameterization is known, all methods we consider are equally effective at high signal-to-noise ratios (SNRs) but marginalization is consistently more robust at low SNRs.
When the parameterization is not known, marginalization over parameterizations has the lowest over-all error rate among the methods we consider.
Furthermore, its error rate is close to that of parameter marginalization given the correct parameterization.
These two cases are examined in Sections \ref{sec:marginalization-over-parameters} and \ref{sec:marginalization-over-parameterizations}.

To assess the relative speed of analytic and numerical marginalization, we examine how two metrics change as the complexity of an inference problem increases: the number of iterations required for MCMC to converge and the number of independent samples generated per unit time.
The basic problem is analyzing a single line on a continuum whose parameterization is known.
To build up more complicated problems, we add more spectral segments each of which has its own continuum and contains another absorption line.
All of these absortion lines share widths and central velocities but have independent column densities.
Problems with this structure arise when analyzing multiple lines from a single species or from multiple species that can be assumed to share a common component structure.
The convergence speedup from using analytic marginalization is dramatic, reaching a full order of magnitude difference in the number of required iterations with as few as three spectral segments.
Analytic marginalization yields more independent samples per unit time when there are multiple spectral segments with high-order continua.
For example, when there are six spectral segments with 3rd order continua, analytic marginalization is three times faster than numerical marginalization.
When there are few spectral segments, analytic marginalization is slightly slower or of comparable speed to numerical marginalization.
These metrics are examined in Section \ref{sec:MCMC-efficiency}.

\subsection{Marginalization over parameters}
\label{sec:marginalization-over-parameters}

\begin{figure}
  \includegraphics[width=\linewidth]{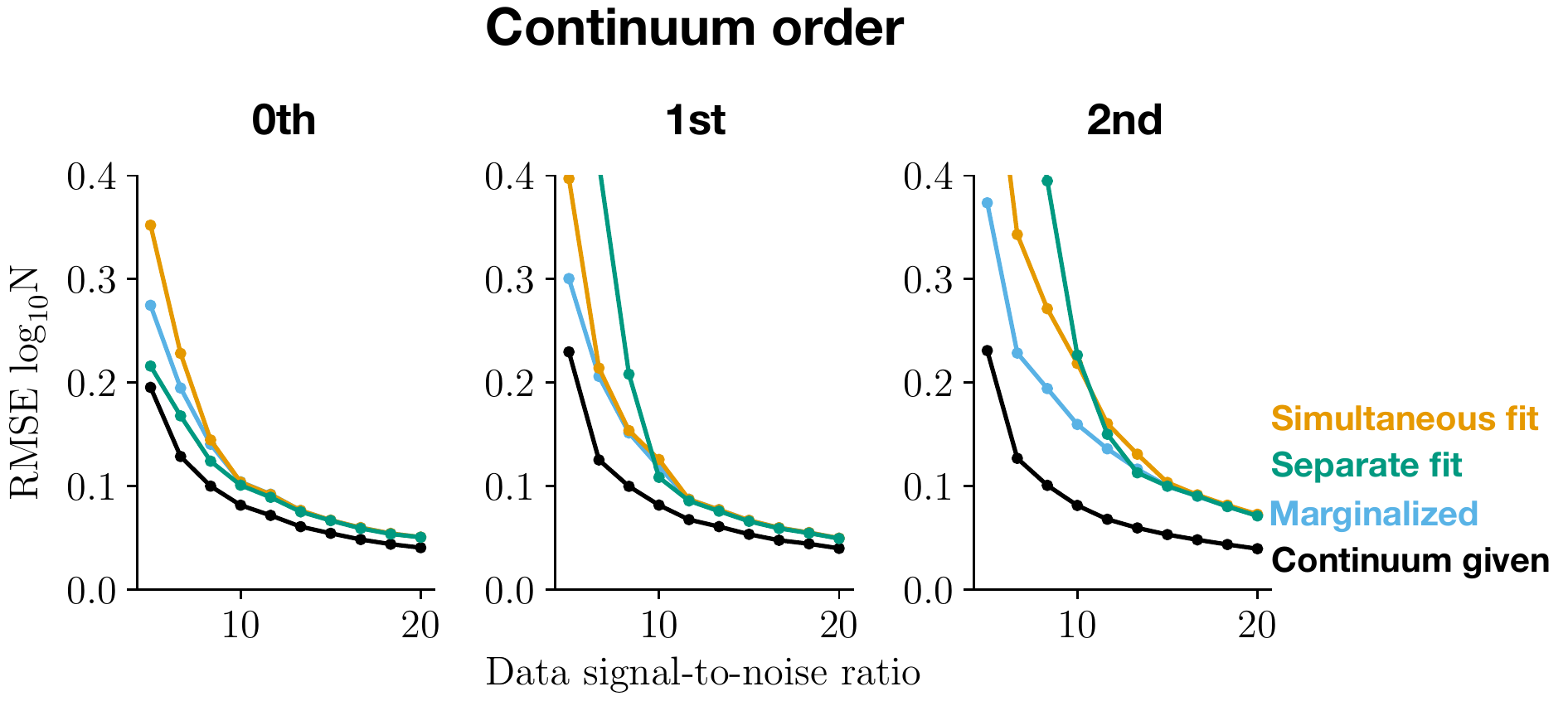}
  \caption{
  Accuracy and precision of different methods of measuring the column density of a single line superimposed on a continuum with known parameterization.
  The accuracy/precision is defined in terms of the root mean square error (RMSE) of the logarithm of the column density measurements.
  The signal-to-noise ratios (SNRs) of the artifically generated spectra used for this test are shown on the x-axis of each panel.
  The panels correspond to different continuum parameterizations, from left to right: 0th order polynomial, 1st order polynomial, 2nd order polynomial.
  The line colors indicate different measurement methods, which are listed in the figure legend.
  These methods are explained in detail in Section \ref{sec:marginalization-over-parameters}.}
  \label{fig:order-known-comparison}
\end{figure}

Here, we consider the problem of measuring the column density of a single well-resolved, unsaturated absorption line superimposed on a continuum whose parameterization is known but whose parameters are not known.
To do this, we generate spectra containing an absorption line with fixed absorption parameters but with varying continuum parameterizations, continuum parameters, and SNRs.
The continuum parameterizations we consider are polynomials of order 0, 1, and 2 and the continuum-level SNRs we use are between 5 and 25.
We generate 1000 spectra at each combination of parameterization and SNR and measure the column density of the absorption line in each spectrum.
From the recovered column densities, we compute the root mean square error (RMSE) of the base ten logarithm of the column density ($\log_{10}{\rm N}$).
We use the logarithm of the column density because its RMSE scale-free.
In particular, physical constants such as oscillator strengths cancel in the $\log_{10}{\rm N}$ RMSE calculation.

We measure the column density using four methods: (1) supply the correct continuum parameters and only fit for the absorption line parameters; (2) simultaneously fit for continuum and absorption line parameters; (3) analytically marginalize over continuum parameters while fitting for absorption line parameters; and (4) use the absorption line parameters recovered using method (1) to define a line-free spectral region, fit continuum parameters just to this region, and with those continuum parameters fit for the absorption line parameters.
The first method provides a lower limit on the RMSE of $\log_{10}{\rm N}$ as a function of SNR.
The second and third methods are two possible ways of automatically modeling the continuum.
The fourth method approximates the actions of a human manually analyzing a spectrum.
We assume the human can correctly estimate the continuum by eye, correctly estimate the best-fit absorption line profile by eye given this continuum, and use this profile to determine which part of the spectrum is not affected by the line.

The RMSEs obtained using these methods are shown in Figure \ref{fig:order-known-comparison}.
Above an SNR of 10-15, all methods where the continuum is not a priori known yield equal results.
At and below that SNR range, marginalization has a lower RMSE than both simultaneous fitting and the human-like analysis.
This advantage becomes greater as the continuum parameterization becomes more complex.

\subsection{Marginalization over parameterizations}
\label{sec:marginalization-over-parameterizations}
\begin{figure}
  \includegraphics[width=\linewidth]{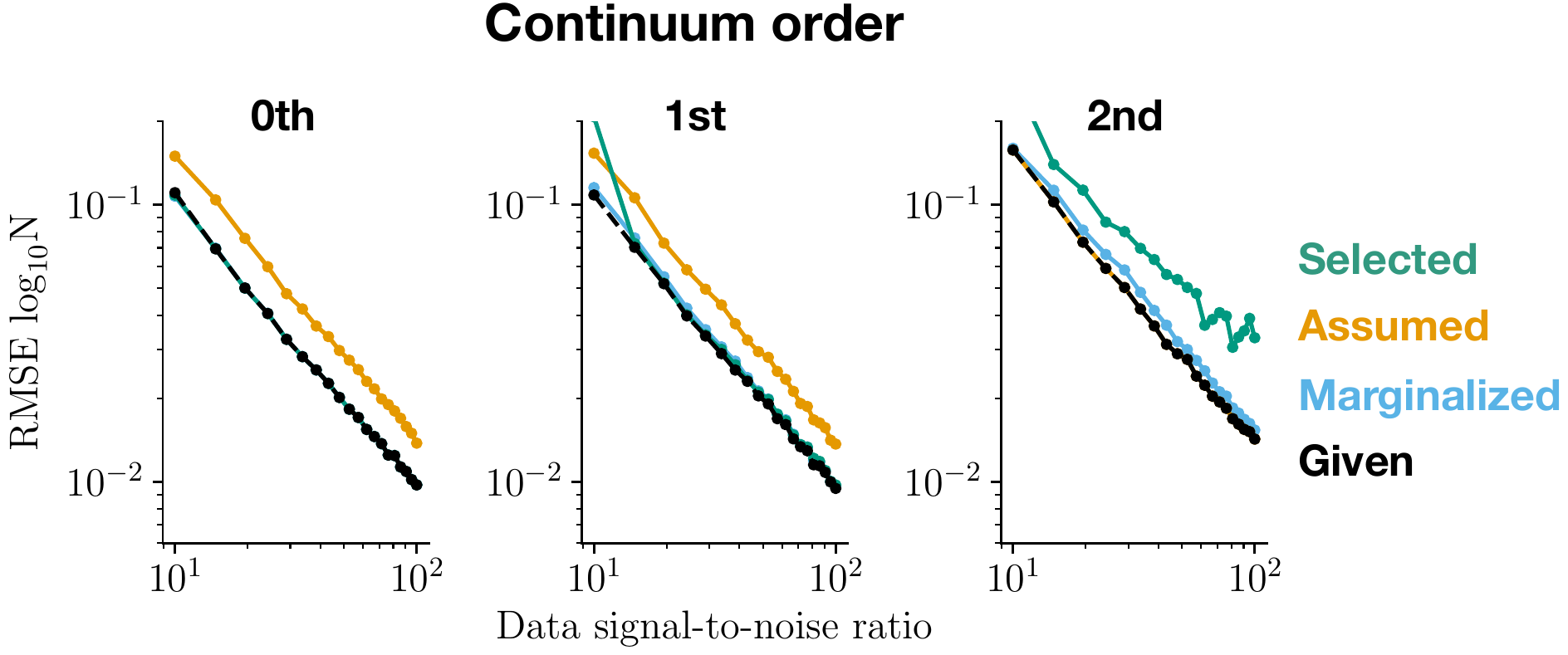}
  \caption{
  Accuracy and precision of different methods of measuring the column density of a single line superimposed on a continuum with unknown parameterization.
  The accuracy/precision is defined in terms of the root mean square error (RMSE) of the logarithm of the column density measurements.
  The signal-to-noise ratios (SNRs) of the artifically generated spectra used for this test are shown on the x-axis of each panel.
  The panels correspond to different true continuum parameterizations, from left to right: 0th order polynomial, 1st order polynomial, 2nd order polynomial.
  The line colors indicate different measurement methods, which are listed in the figure legend.
  These methods are explained in detail in Section \ref{sec:marginalization-over-parameterizations}.
  }
  \label{fig:order-unknown-comparison}
\end{figure}
Next, we consider a problem where there is still a single well-resolved and unsaturated absorption line but where it is only known that the continuum belongs to a \emph{family} of possible continuum parameterizations.
As in the previous section, we consider three possible continuum parameterizations: 0th, 1st, and 2nd order polynomials.
The approach, simulating spectra, measuring $\log_{10}N$ for each simulated spectrum, and computing the RMSE of $\log_{10}N$, is also the same.
However, we consider a different range in continuum-level SNR: 10 to 100.

We measure the column density using four methods: (1) supply the correct parameterization and marginalize over its parameters; (2) assume the most complicated of the three parameterizations and marginalize over its parameters; (3) use an iterated likelihood ratio test to choose a parameterization and fit for its parameters; (4) marginalize over parameterizations as well as parameters.
Method (1) establishes a reference minimum RMSE for this test case.
Method (2) is a conservative assumption that can be made when the family of possible parameterizations is nested---a polynomial of order $n$ with leading coefficient 0 is a polynomial of order $n-1$.
Methods (3) and (4) are different ways of automatically accounting for the different possible parameterizations, in one case (3) by selecting a parameterization and in the other (4) by averaging over the possible parameterizations.
The likelihood function that we maximize when using method (4) is the weighted sum of the continuum-marginalized likelihoods of the three continuum models.
The weights in this sum are the prior probabilities of each of the models; we assume all three are equally likely.

The RMSEs of the four methods are shown in Figure \ref{fig:order-unknown-comparison}.
We compare the methods using three criteria: their robustness to decreasing the SNR and increasing the number of parameters relative to the number of observations; the value of their RMSE at fixed SNR; and the SNR they require to obtain the same RMSE.
We consider a method to be robust if its RMSE scales consistently with SNR and true continuum order rather than rapidly increasing at some critical value.
The reference, conservative, and parameterization-marginalization methods are robust for SNRs between 10 and 100 while the parameterization-selection method is not.
Its RMSE blows up below an SNR of 10 for spectra with 1st order continua and at all SNRs considered for 2nd order continua.
While parameterization selection performs as well as the other methods on high SNR spectra, its lack of robustness means that it is not as generally applicable as the other methods.

In terms of RMSE, the parameterization-marginalized estimator is nearly as good as the reference estimator.
The ratio RMSE$\rmsub{m}$/RMSE$\rmsub{r}$ of the parameterization-marginalized RMSE, RMSE$\rmsub{m}$, to the reference RMSE, RMSE$\rmsub{r}$, is 1, 1.04, and 1.1 for spectra with 0th, 1st, and 2nd order continua.
For spectra with 0th or 1st order continua, the conservative estimator is significantly worse than the reference estimator.
RMSE$\rmsub{c}$/RMSE$\rmsub{r}$ is 1.5, 1.5, and 1, respectively, where RMSE$\rmsub{c}$ is the RMSE of the conservative estimator.
These ratios are approximately constant across the entire considered SNR range.
When analyzing an already acquired set of observations in which a variety of continuum parameterizations are present, using the parameterization-marginalization estimator rather than the conservative estimator will, on average, yield higher accuracy and precision.

It is also useful to compare the SNRs the conservative and parameterization-marginalization estimators require to achieve the same RMSE.
For spectra with 0th, 1st, and 2nd continua, the ratio of the required SNRs SNR$\rmsub{c}$/SNR$\rmsub{m}$ is 1.6, 1.5, and 0.9.
These ratios are, again, consistent across the entire considered SNR range.
Assuming that SNR is proportional to the square root of observing time, as is the case for Poisson noise-limited data, these SNR ratios can be converted to required observing time ratios.
Reaching the RMSE of the parameterization-marginalized estimator with the conservative estimator takes 1.26, 1.22, and 0.95 times as much observing time.
When designing an observing strategy to meet a column density RMSE requirement, using the parameterization-marginalized estimator rather than the conservative estimator can save observing time given a fixed sample or increase the size of a sample given a fixed amount of observing time.

\subsection{MCMC efficiency}
\label{sec:MCMC-efficiency}

\begin{figure}
  \includegraphics{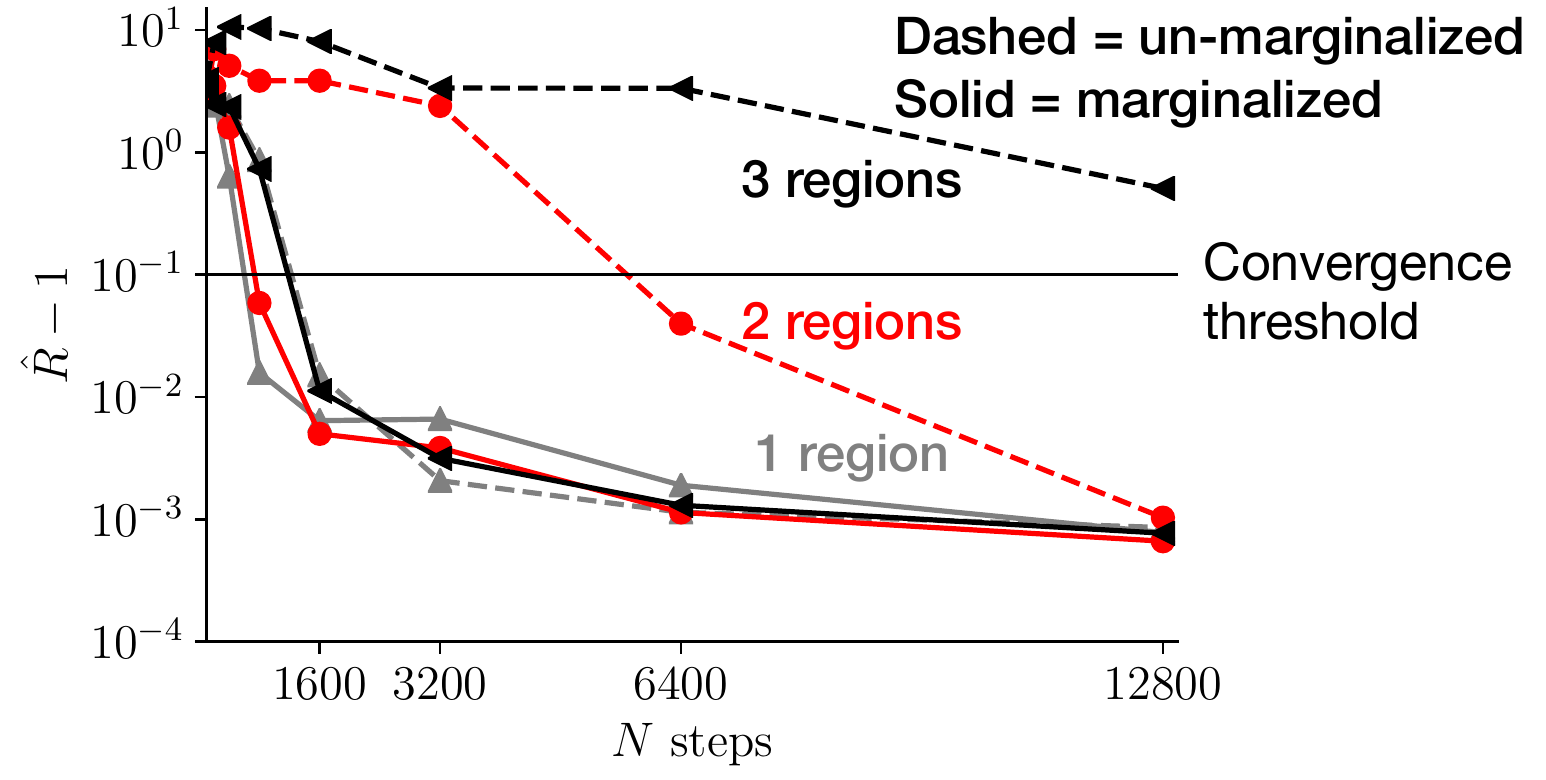}
  \caption{
  Convergence rate of MCMC with analytic and numerical continuum parameter marginalization for absorption line analysis problems with different complexities.
  The convergence diagnostic (y-axis) is the Rubin-Gelman statistic, an estimate of how much smaller the Monte Carlo error of an MCMC-based parameter estimate can get.
  Each line shows the evolution of this convergence diagnostic as a function of the number of MCMC steps taken (x-axis).
  Line styles indicates whether continuum parameters are marginalized over analytically (solid) or included in MCMC (dashed).
  Line colors and markers indicate the number of spectral regions being analyzed simultaneously; each region has its own set of continuum parameters.
  The Rubin-Gelman statistic and the problem setup are discussed in more detail in Section \ref{sec:MCMC-efficiency}.
  }
  \label{fig:convergence-comparison}
\end{figure}

\begin{figure}
  \includegraphics{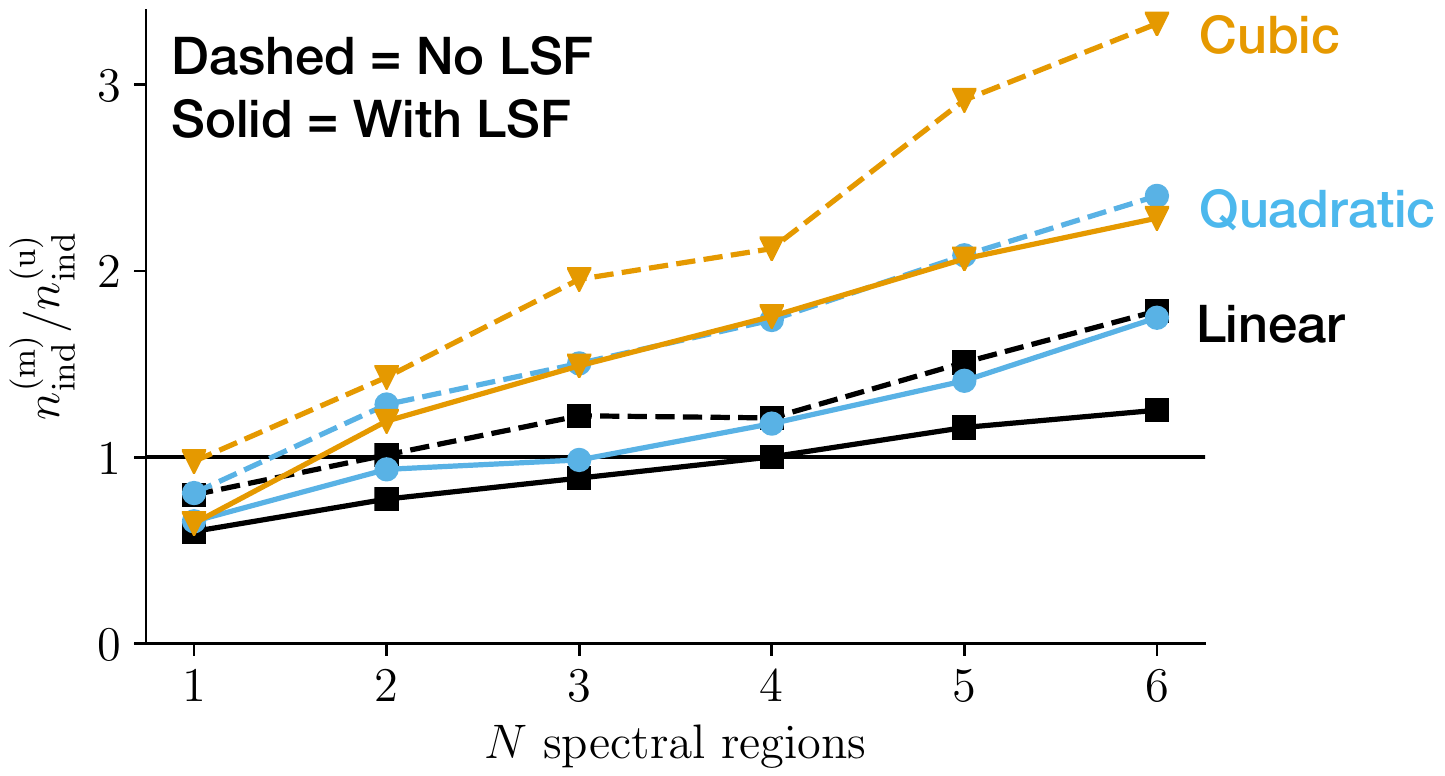}
  \caption{
  Relative efficiency of MCMC with analytic and numerical continuum parameter marginalization for absorption line analysis problems with different complexities.
  The relative efficiency is the ratio of the number of independent samples, $n_{\rm ind}$, generated in the same amount of time by the two marginalization approaches; $n_{\rm ind}^{(m)}$ uses the analytically marginalized likelihood, $n_{\rm ind}^{(u)}$ uses the unmarginalized likelihood.
  The larger the relative efficiency, the more independent samples generated by analytic marginalization.
  Line colors and markers correspond to different continuum parameterizations: 1st order polynomial (black squares), 2nd order polynomial (blue circles), 3rd order polynomial (orange triangles).
  Line styles indicate whether a non-trivial LSF is used in the analysis.
  The relative efficiency is shown as a function of the number of spectral regions being analyzed simultaneously; each spectral region has its own set of continuum parameters.
  The relative efficiency and the problem setup are discussed in more detail in Section \ref{sec:MCMC-efficiency}.
  }
  \label{fig:efficiency-comparison}
\end{figure}

In ISM absorption spectra, it is common to have multiple lines in a spectrum with shared parameters.
These lines can be from the same species, e.g. the Lyman series, or from different species, e.g. from Mg {\small I}, Zn {\small II}, and Cr {\small II} which have overlapping lines in the near ultraviolet.
When these lines are in different parts of a spectrum, each part needs its own continuum parameters.
This is a case in which analytic marginalization can potentially be more efficient than MCMC marginalization.

We compare how quickly MCMC done using each of the two methods converges and how efficient MCMC done using each method is post-convergence.
Which comparison is more informative for choosing a method to use will depend on the purpose of the MCMC run.
If the goal of an MCMC run is to estimate some value at low-to-moderate precision, the rate of convergence will be the more important factor.
If the goal is instead to estimate some value at high precision, the burn-in period will usually be a small fraction of the total chain and post-convergency efficiency will be more important.

We consider a case where there are $N$ absorption lines with shared central velocities and widths and independent column densities.
Each absorption line is in a different spectral region.
The continuum in each spectral region is a polynomial of order $M$.
The marginalized likelihood has $2 + N$ absorption line parameters.
The unmarginalized likelihood has $2 + N$ absorption line parameters and $N \times M$ continuum parameters.
We use the \texttt{emcee} implementation of the Goodman and Weare affine-invariant MCMC ensemble sampler to generate draws from the posterior corresponding to each of these likelihoods.
We use the minimum number of ``walkers,'' which is twice the number of parameters.

We use the Rubin-Gelman statistic $\hat{R}$ \citep{Gelman:1992zz} to assess convergence.
The Rubin-Gelman statistic compares the variance between and within different MCMC instances.
If the instances have all converged, these two variances should be approximately equal.
We run ten MCMC instances for 12800 (per-walker) steps and compute the Rubin-Gelman statistic from the second half of sub-chains of length $2^p \times 100$ for $p=0, 1, \ldots, 7$.
$\hat{R}$ is computed separately for each parameter.
Following common usage, we consider convergence to be reached when the $\hat{R}$ of all parameters is less than $1.1$.
We run this test for 1, 2, and 3 regions and absorption lines assuming a continuum of order 1.
The value of the $\hat{R}$ as a function of (total) number of steps is shown in Figure \ref{fig:convergence-comparison}.
When there is a single region and line, the MCMC marginalization chain takes twice as many steps as the analytic marginalization chain to converge; when there are two regions, it takes eight times as many steps; when there are three, the MCMC marginalization chain has not converged by the maximum chain length of 12800 while the analytic marginalization chain converges within 1600 steps.

We use the number of independent samples per unit time to assess efficiency.
We run MCMC with the marginalized likelihood for 2000 burn-in steps and 8000 converged steps and record the average time per sample, $t_s$.
Because MCMC with the unmarginalized likelihood takes many steps to converge, we use draws from the converged part of the marginalized likelihood chain as a starting point; these draws only have values for the absorption line parameters.
At each set of absorption line parameters, we sample a set of continuum parameters from the conditional distribution discussed in Section \ref{sec:conditionals}.
From this starting point, we run MCMC with the unmarginalized likelihood for 4000 burn-in steps and 36000 converged steps and record the average time per sample.
We then compute the average integrated autocorrelation times $\tau_f$ of the walkers in both chains.
The number of independent samples per unit time is $n_i = \left(\tau_f \, t_s \right)^{-1}$.

We compute $n_i$ for a number of regions $N = 1, 2, \ldots, 6$, continua of polynomial order $M=$ 1, 2, and 3, and either a trivial LSF or a banded LSF.
The ratio $n_{ind}^{(\text{m})} / n_{ind}^{(\text{u})}$ for each of these cases is shown in Figure \ref{fig:efficiency-comparison}.
When this ratio is greater than 1, running MCMC with the marginalized likelihood for a fixed amount of time will produce more independent samples than running MCMC with the unmarginalized likelihood for the same amount of time.
The greater the number of regions and the order of the continuum, the greater the efficiency advantage of the marginalized likelihood over the unmarginalized likelihood.
This advantage will not depend on the number of datapoints in each spectral region so long as the LSF is trivial or banded, since in these cases the evaluation time of both likelihoods grows linearly with dataset length (see Section \ref{sec:scaling}).

\section{Discussion}
\label{sec:discussion}
\subsection{Assumptions and consequences}
The explicit assumptions of the analytic marginalization method are that the continuum is a linear function, that the prior on the coefficients is the improper uniform or multivariate normal distribution, that residuals between the data and model are normally distributed, and that the covariance matrix of the residuals does not depend on the continuum.
It is obvious that these assumptions do not hold strictly for any dataset.
For example, both possible priors require that there not be constraints on the coefficients even though no background source produces negative flux.
A less trivial example is data in the low photon count regime, which are better described by a Poisson distribution than a Gaussian distribution.
This is particularly important when the uncertainties on the measurements are themselves highly uncertain and should be explicitly modeled.
In that case, the uncertainties will depend on the Poisson intensity function, which explicitly depends on the continuum.
Analytic marginalization of the kind described in this work should not be applied to low SNR X-ray or UV spectra.

An implicit assumption of the method is that the absorption model is realistic.
For analytic marginalization to be useful, it must be possible for the absorption model to correctly describe the actually present absorption features.
For example, if a region of a spectrum contains two clearly distinct absorption lines but the model only allows for a single line, the presence of the un-modeled line will bias the continuum model.
In short, improvements in continuum modeling cannot solve problems of absorption model misspecification.

The continuum models envisaged in this work will usually be effective descriptions rather than (often non-linear) physical descriptions.
Most continua that vary over longer wavelength scales than the width of absorption lines in question can be approximated in this way.
Examples of background sources with slowly varying continua include quasars and (particularly rapidly rotating) hot stars.
With flexible linear models such as splines, it is even possible to describe more complicated pseudo-continua such as stellar wind lines.
For even more complicated pseudo-continua such as those of cool stars \citep[e.g.]{Zasowski:2015hi}, it is necessary to use a non-linear model.
Marginalizable linear models can still be useful even in this case as a way of introducing small corrections for pseudo-continuum features that are not perfectly described by the non-linear model.

\subsection{Applications of analytic marginalization}
The test cases in Section \ref{sec:test-cases} showed that marginalization over continuum parameters and parameterizations is more precise, accurate, and robust than the alternatives.
Considered purely as a replacement for numerical marginalization, analytic marginalization is just a potentially more computationally efficient way of implementing an existing inference approach.
However, it also allows two qualitatively new approaches: continuum model averaging and absorption parameter optimization with a continuum-marginalized likelihood function.

The test case in Section \ref{sec:marginalization-over-parameterizations} combines both of these approaches---optimizing an absorption parameter likelihood function where the parameterization and parameters of the continuum have been marginalized over.
Analytic marginalization makes this possible in two ways: availability of closed form likelihoods and availability of gradients of closed form likelihoods.
Optimization with continuum-marginalized likelihoods is useful for analyzing large surveys.
Analyses of absorption lines in tens of thousands of spectra \citep[e.g.]{2013ApJ...770..130Z,Zasowski:2015hi} cannot practically be done with MCMC.
With analytic marginalization, it is possible to at least marginalize over continuum parameters.
The results of the test cases suggest that this approach could mean a non-trivial improvement in the accuracy and precision of absorption line measurements.

In cases where MCMC is possible, combining continuum parameterization marginalization with a probabilistic specification of absorption component structure would allow absorption line analysis with human intervention only at the level of specifying priors and candidate continuum parameterizations.
Component structure specification can be done using trans-dimensional inference, in which the dimensionality of parameter space (in this case the number of sets of absorption line parameters) is itself a parameter of the model.
This way of doing absorption line analysis has two potential advantages.
First, marginalizing over the velocity structure of the absorption as well as the continuum should automatically include effects such as unresolved saturated structure in parameter uncertainties.
Second, because inference approach is almost completely automatic, it allows blinding, which improves reproducibility.

\section{Conclusion}
\label{sec:conclusion}
Absorption lines are an important source of information about stars and the ISM.
As larger spectroscopic datasets become available and as reproducibility becomes more standard in astronomy, it becomes necessary to move beyond ad-hoc analysis methods, particularly ones in which a human directly interacts with data.
In multiple recent works, there have been attempts to partially automate continuum placement by including and marginalizing over continuum parameters in probabilistic spectral models.
Marginalizing over continuum parameters has, in these works, been hypothesized to also improve the accuracy of the recovered absorption line parameters.
Despite these advantages, this approach has so far not become popular, in part due to the computational expense of numerically marginalizing over these additional parameters.

In this work, we have shown that it is possible in many cases to replace this numerical marginalization with analytic marginalization (Section \ref{sec:assumptions-and-formalism}).
Analytic marginalization speeds up MCMC-based analyses in problems with many continuum parameters (Section \ref{sec:MCMC-efficiency}).
The continuum parameter-marginalized likelihood can also be used for optimization over absorption line parameters.
This approach combines the speed of optimization with the advantages of continuum marginalization.
Analytic marginalization over continuum parameters makes it trivial to also marginalize over continuum \emph{parameterizations}.
As with parameter marginalizaton, parameterization marginalization can be combined with optimization over absorption line parameters.
parameterization marginalization further reduces the amount of direct human interference in the analysis of individual spectra and will be especially useful in analyses of datasets containing spectra with different continuum shapes.

We have also confirmed that marginalization over continuum parameters and parameterizations indeed improves the accuracy of absorption line parameter measurements.
The advantage of parameter marignalization is only significant at low SNRs (Section \ref{sec:marginalization-over-parameters}).
On the other hand, parameterization marginalization is significantly more accurate than alternative methods of deciding on a continuum parameterization at all SNRs (Section \ref{sec:marginalization-over-parameterizations}).

We have released an open-source \texttt{python} package, \pkgname, which can be used to evaluate continuum parameter-marginalized likelihoods and related quantities.
Features of this package are described in Appendix \ref{sec:package-and-demos}.
It is meant to be used as a drop-in replacement for likelihood functions in existing absorption spectrum analysis tools.

\acknowledgments
The author thanks Andrew Casey, Andrew Fox, Cameron Liang, and Yong Zheng for useful discussions about use cases and Joshua Peek and Linda Tchernyshyov for helpful comments.

\software{emcee \citep{2013PASP..125..306F},
matplotlib \citep{2007CSE.....9...90H},
numpy \citep{vanderWalt:dp},
scipy \citep{SciPy}
}

\appendix

\section{Implementation and demonstration}
\label{sec:package-and-demos}
In this Appendix, we describe how \pkgname{} is implemented (Section \ref{sec:implementation}), list some of its capabilities (Section \ref{sec:package-functionality}), and show how the computation time of different calculations grows with dataset and continuum model size (Section \ref{sec:scaling}).

\subsection{Implementation}
\label{sec:implementation}
We have implemented \pkgname{} as a pure-Python package with \texttt{numpy} and \texttt{scipy} as dependencies.
\pkgname{} does not contain functionality for building LSFs or computing transmittances from absorption parameters and is not intended to be a stand-alone analysis tool.
It is meant to be used as a drop-in likelihood function replacement in analysis packages or scripts.

\subsection{Package functionality}
\label{sec:package-functionality}
This package was designed for a use case where the log marginal likelihood and its gradient are evaluated at many different values of the $\theta$-dependent parameters (see Section \ref{sec:assumptions-and-formalism}) while the $\theta$-independent parameters are held constant.
The core feature of the package is the \texttt{MarginalizedLikelihood} class.
A \texttt{MarginalizedLikelihood} instance stores $\theta$-independent parts of the model and pre-computes quantities that are re-used during repeated marginalized likelihood evaluations.
In particular, it stores the data covariance matrix $\vx{K}$; the $\vx{c}$ prior covariance matrix $\vx{\Lambda}$ and its explicit inverse, if applicable; and the LSF mapping $\vx{L}$ and its transpose.

Both covariance matrices can be diagonal or fully general.
The package includes the \texttt{CovarianceMatrix} class, which defines a consistent interface for calculations, and two subclasses, \texttt{DiagonalCovarianceMatrix} and \texttt{GeneralCovarianceMatrix}.
\texttt{DiagonalCovarianceMatrix} wraps the simple, one-dimensional determinant and inverse calculations possible with a covariance matrix consisting purely of variances and does the book-keeping required to produce output with the correct shape.
\texttt{GeneralCovarianceMatrix} uses the Cholesky decomposition of the supplied covariance matrix to calculate its determinant and to left multipy matrices and vectors by its inverse.
Computing the Cholesky decomposition of a general covariance matrix of size $M$ by $M$ takes $\mathcal{O}(M^3)$ calculations, making it prohibitively computationally expensive for large $M$.

The LSF mapping $\vx{L}$ can be any object that implements the matrix multiplication interface, i.e. has a \texttt{matmul} or \texttt{\_\_matmul\_\_} method.
For example, $\vx{L}$ can be a dense matrix represented by a \texttt{numpy} array, a sparse matrix represented by a \texttt{scipy.sparse} matrix, or a convolution operator represented by a \texttt{scipy.sparse.linalg} \texttt{LinearOperator}.
$\vx{L}$ can also be the identity mapping (indicated by \texttt{None}), in which case it is left out of any likelihood calculations.

\subsection{Computation time as a function of dataset and basis size}
\label{sec:scaling}
The most time-consuming step in computing all of the quantities derived in Section \ref{sec:assumptions-and-formalism} is forming the matrix $\vx{C}_{n/u}$.
This step requires matrix-matrix products, while most other steps only involve matrix-vector products.
These expensive products are $\vx{L}\vx{B}$ and $\vx{K}^{-1} \left(\vx{L} \vx{B}\right)$.
The amount of time required to compute these products depends on the structure $\vx{L}$ and $\vx{K}$.

$\vx{L}$ can be the identity matrix, a dense matrix, a sparse matrix, or a linear mapping such as convolution.
The fastest case is when $\vx{L}$ is the identity matrix, since then $\vx{L}\vx{B}$ does not need to be computed.
The slowest case is when it is a dense matrix, in which case computation time grows as $\mathcal{O}(MN(P+Q))$.
When $\vx{L}$ is a sparse matrix or linear mapping, the scaling depends on its exact structure.
An LSF that varies with wavelength can be represented by a banded matrix, which will be sparse if the spectrum spans many resolution elements.
If the bandwidth of $\vx{L}$ is independent of the size of the dataset, the computation time of this product grows as $\mathcal{O}(M(P+Q))$.

We consider covariance matrices $\vx{K}$ that are either diagonal or general.
If $\vx{K}$ is diagonal, $\vx{K}^{-1} \left(\vx{L} \vx{B}\right)$ requires exactly $M(P+Q)$ multiplications.
When $\vx{K}$ is a general covariance matrix, we decompose it into its Cholesky factors and left-multiply $\vx{L} \vx{B}$ by $\vx{K}^{-1}$ by solving the linear problem $\vx{\vx{L} \vx{B}} = \vx{K} \vx{X}$.
The time needed to factor $\vx{K}$ grows as $\mathcal{O}\left(M^3\right)$ but only needs to be done once per set of observations.
The time needed to solve the linear problem grows as $\mathcal{O}\left(M^2 (P+Q)\right)$.

\begin{figure}
  \includegraphics{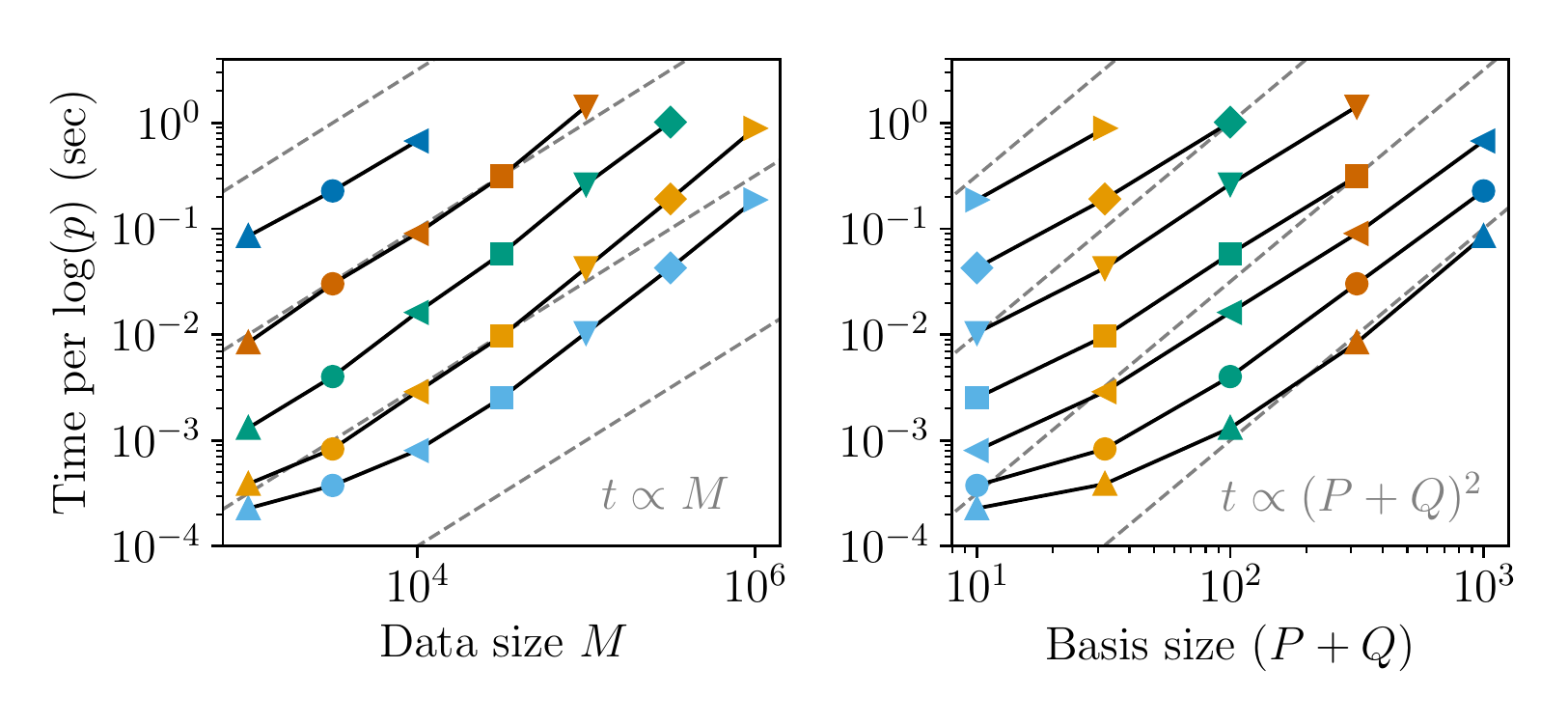}
  \caption{Computation time of the marginal log-likelihood (Equations \ref{eqn:proper-prior-marginal} and \ref{eqn:improper-prior-marginal}) when the data covariance matrix $\vx{K}$ is diagonal and $\vx{L}$ is the identity mapping as a function of dataset size $M$ (left panel) and basis size $P+Q$ (right panel). Values with the same marker shape were computed at the same dataset size $M$. Values with the same marker color were computed at the same dataset size $P+Q$. Polynomials of the form given in the bottom right corner of each panel are shown as dashed gray lines.}
  \label{fig:uncorr-no-L-logp}
\end{figure}

\begin{figure}
  \includegraphics{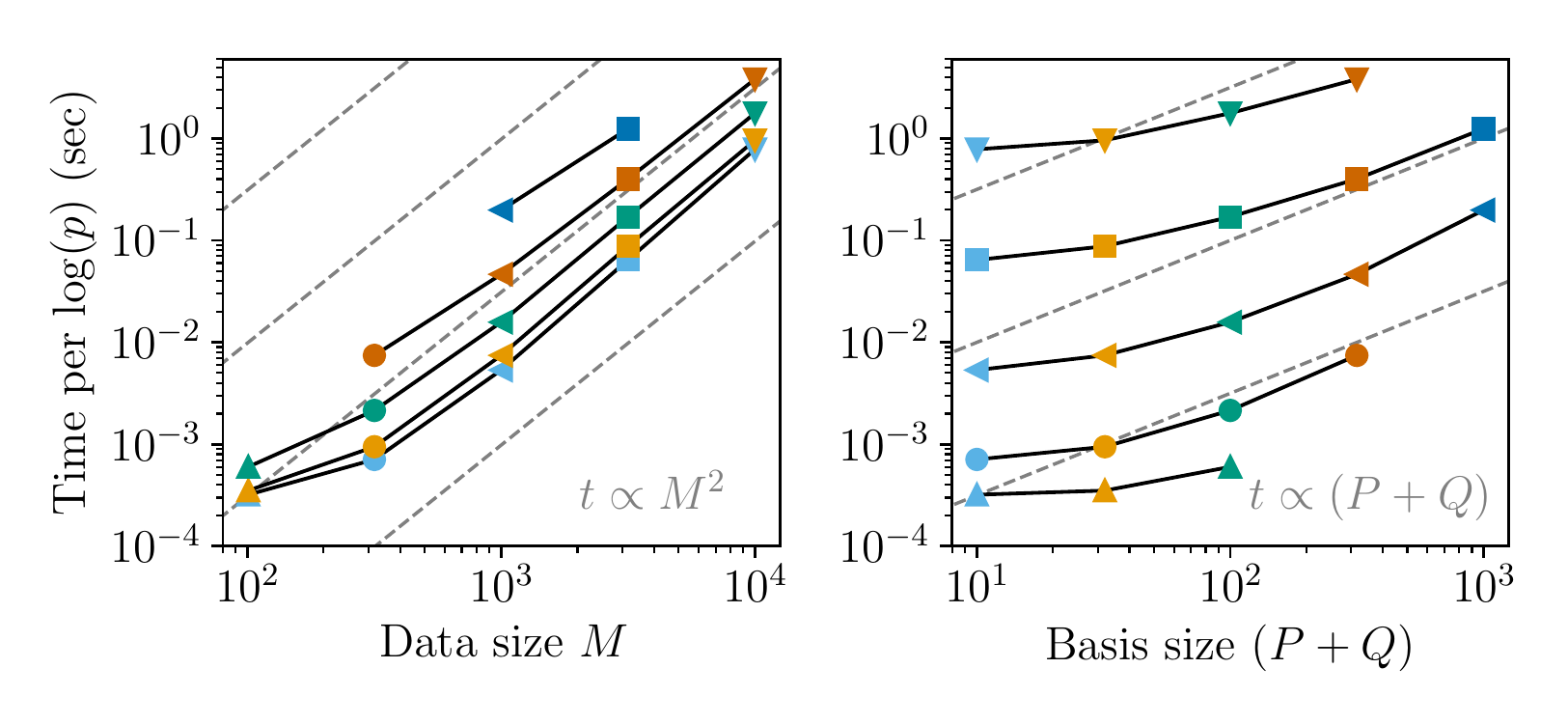}
  \caption{Computation time of the marginal log-likelihood when the data covariance matrix $\vx{K}$ is not diagonal and $\vx{L}$ is a dense matrix. See caption of Figure \ref{fig:uncorr-no-L-logp} for a description of figure elements.}
  \label{fig:corr-yes-L-logp}
\end{figure}

\begin{figure}
  \includegraphics{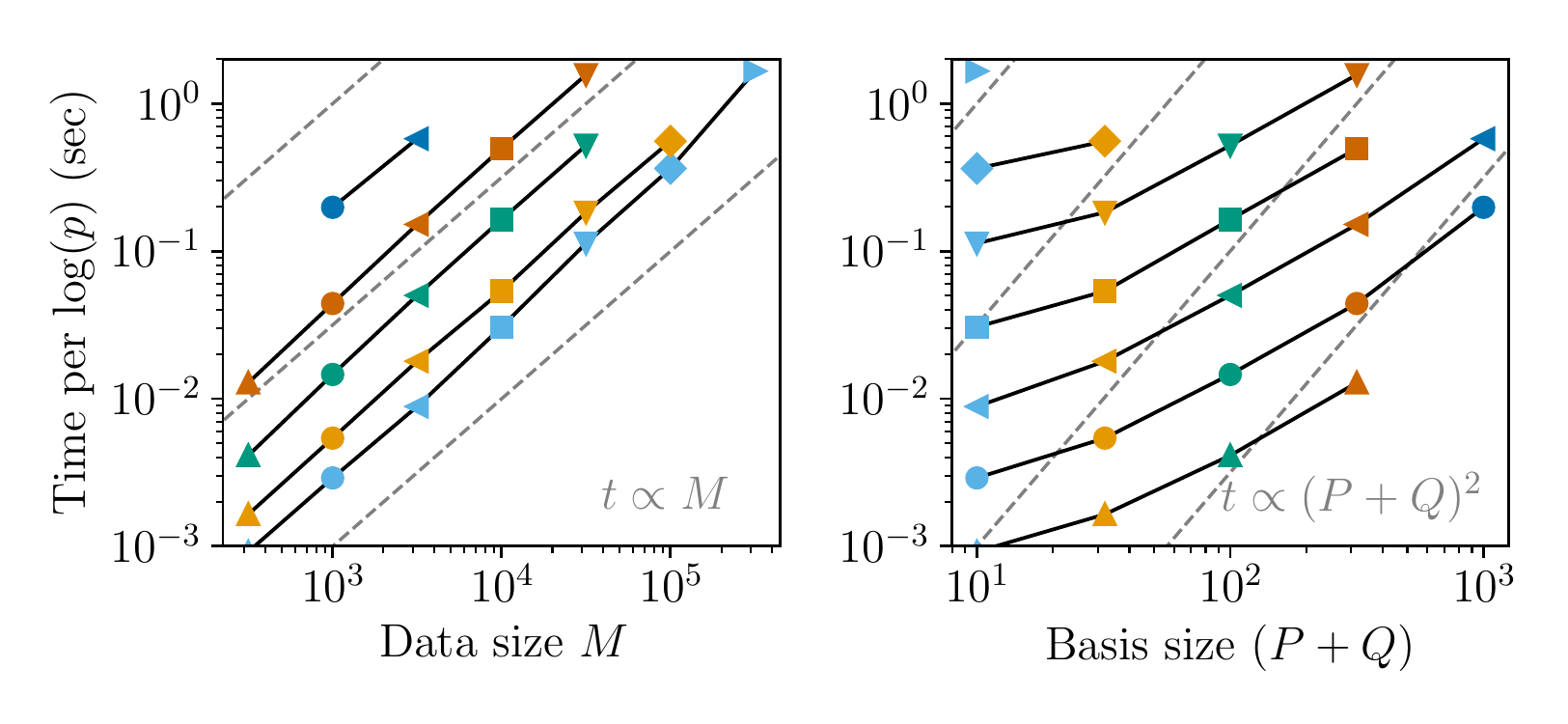}
  \caption{Computation time of the marginal log-likelihood when the data covariance matrix $\vx{K}$ is diagonal and $\vx{L}$ is a sparse, banded matrix. See caption of Figure \ref{fig:uncorr-no-L-logp} for a description of figure elements.}
  \label{fig:uncorr-sparse-L-logp}
\end{figure}

To empirically confirm these growth rates, we timed how long it takes to evaluate the log-likelihood and its gradient for a range of dataset sizes $M$ and basis sizes $P+Q$ and three $\vx{L}$ and $\vx{K}$ structure scenarios.
The scenarios are: $\vx{L}$ is the identity mapping, $\vx{K}$ is diagonal; $\vx{L}$ is a dense matrix, $\vx{K}$ is general; and $\vx{L}$ is a sparse, banded matrix and $\vx{K}$ is diagonal.
The first two scenarios are the fastest and slowest combination.
The third scenario is more typical for a spectrum; the data uncertainty is diagonal and the LSF has finite extent.
The evaluation time of the log-likelihood as a function of $M$ and $P+Q$ for these three scenarios is shown in Figures \ref{fig:uncorr-no-L-logp}, \ref{fig:corr-yes-L-logp}, and \ref{fig:uncorr-sparse-L-logp}.
We do not show the evaluation time of the gradient because it behaves in the same way as the evaluation time of the log-likelihood in all three scenarios; the most expensive step of the two calculations is the same.

The dependence of computation time on $M$ and $P+Q$ generally agrees with the predictions based on the two most time-consuming steps.
At low $M$ and in particular at low $P+Q$, the computation time is either overhead-dominated or evenly split between the most time-consuming steps and other steps.
When $M \gtrsim 10^5$, computation time increases faster than expected purely from the growth rate of the required number of operations (see e.g. the left panel of Figure \ref{fig:uncorr-no-L-logp}).
This excess increase in computation time is most likely due to changes in memory bandwidth, as the size of matrix rows and columns increases past the size of the highest-level CPU cache on the laptop used to run these tests.

To put these dataset sizes into context, a Sloan Digital Sky Survey (SDSS) BOSS or APOGEE spectrum is $\sim 10^3$ pixels long, a Hubble Space Telescope Cosmic Origins Spectrograph (HST-COS) spectrum is $\sim 10^4$ pixels long, and a spectrum from an echelle spectrograph such as the Ultraviolet and Visual Echelle Spectrograph on the Very Large Telescope or the Magellan Inamori Kyocera Echelle spectrograph is $\sim 10^5 - 10^6$ pixels long.
The uncertainties associated with these spectra are usually assumed to be diagonal and the LSFs are acceptably described by sparse, banded matrices, so the computation times given in Figure \ref{fig:uncorr-no-L-logp} and \ref{fig:uncorr-sparse-L-logp} should apply.

\end{document}